\documentclass[letterpaper,11pt]{article}

\usepackage{dynkin-diagrams}
\usepackage{cite}
\usepackage{amsmath}
\usepackage{amssymb}
\usepackage{amsthm}
\usepackage{mathrsfs}
\usepackage{xypic}
\usepackage{tikz-cd}
\usepackage{chngcntr}
\usepackage{enumitem}
\usepackage{cancel}
\usepackage{authblk}
\usepackage[normalem]{ulem}

\usepackage{hyperref}
\usepackage{cleveref}
\usepackage{color}
\definecolor{darkred}{rgb}{0.65,0.15,0}
\hypersetup{pdfborder={0 0 0},colorlinks=true,urlcolor=darkred,citecolor=blue,linkcolor=darkred,linktocpage=true}

\usepackage{float} 

\usepackage{geometry}

\newgeometry{vmargin={35mm}, hmargin={35mm}} 

\usepackage{youngtab}
\def\youngdim{8pt}
\def\youngexpdim{5pt}
\Yboxdim\youngdim

\usepackage{caption}

\def\xyng{\Yboxdim\youngexpdim\yng(1)\Yboxdim\youngdim}

\def\eg{{\it e.g.}}
\def\ie{{\it i.e.}}

\def\DWeight#1#2#3{\bigl(\raise2.5pt\hbox{${}_{#1}$}{}^{#2}_{#3}\bigr)}

\def\AAWeight#1#2{\bigl(\raise0pt\hbox{${}^{#1}_{#2}$}\bigr)}

\def\fg{{\mathfrak g}}
\def\fh{{\mathfrak h}}

\def\so{{\mathfrak{so}}}

\def\db{\textbf{d}}

\def\nn{\nonumber}

\def\so{\mathfrak{so}}

\def\*{\partial}

\def\RR{{\mathbb R}}
\def\CC{{\mathbb C}}
\def\ZZ{{\mathbb Z}}

\def\NC{{\mathscr N}}

\def\mA{{\mathscr A}}
\def\mI{{\mathscr I}}

\def\Log{\hbox{Log}}
\def\Exp{\hbox{Exp}}

\def\tint{{\textstyle\int}}

\newlength\symlength
\newlength\pluslength
\makeatletter
\DeclareRobustCommand{\cev}[1]{%
  {\mathpalette\do@cev{#1}}%
}
\newcommand{\do@cev}[2]{%
  \vbox{\offinterlineskip
    \sbox\z@{$\m@th#1 x$}%
    \ialign{##\cr
      \hidewidth\reflectbox{$\m@th#1\vec{}\mkern4mu$}\hidewidth\cr
      \noalign{\kern-\ht\z@}
      $\m@th#1#2$\cr
    }%
  }%
}
\makeatother

\def\Cev#1{\overset\leftarrow{#1}}
\def\Vec#1{\overset\rightarrow{#1}}

\def\Sym{\mathrm{Sym}}

\makeatletter
\newcommand{\oset}[3][0ex]{%
  \mathrel{\mathop{#3}\limits^{
    \vbox to#1{\kern-3\ex@
    \hbox{$\scriptscriptstyle#2$}\vss}}}}
\makeatother

\def\Str{\hbox{Str}}

\def\expyng#1{\Yboxdim\youngexpdim\yng(#1)\Yboxdim\youngdim}

\title{Classical BV cohomology of the $N=1$ spinning particle\vspace{18pt}}

\author[1]{Eugenia Boffo\thanks{boffo@karlin.mff.cuni.cz }}
\author[2]{Martin Cederwall\thanks{martin.cederwall@chalmers.se}}

\affil[1]{\small{Faculty of Mathematics and Physics, Mathematical Institute, Charles University Prague, Sokolovsk\'a 83, 186 75 Prague, Czech Republic}\vspace{6pt}}
\affil[2]{\small{Dept. of Physics, Chalmers University of Technology, 412 96 Gothenburg, Sweden}\vspace{6pt}}

\date{\vspace{-5ex}}

\numberwithin{equation}{section}

\begin{document}

    \frenchspacing

\maketitle

\begin{abstract}
\noindent We show that the classical Batalin--Vilkovisky cohomology at negative ghost number of the spinning particle, observed in ref. \cite{Getzler:2015jrr}, is removed by a Koszul--Tate resolution involving saturation of Grassmann odd variables. The model thus satisfies the axioms of Felder and Kazhdan. 
The AKSZ formulation of the resolved model is described.
We reveal partial information on the resolution of the constrained phase space, which involves resolving the parity-shifted tangent sheaf of the light-c\^one. Specialising to dimension one, we describe the full resolution. 
\end{abstract}

\newpage

\tableofcontents

\section{Introduction and summary}

The Batalin--Vilkovisky (BV) formalism \cite{Batalin:1981jr} plays a pivotal r\^{o}le for the perturbative quantisation of gauge theories, especially when the gauge algebra does not close unless the field equations are used. 
A field theory with action functional $S_0$ is cast in the classical BV setup when a new functional $S$, which is defined on an extended space of fields that includes the original gauge fields and resolves the quotient by gauge symmetries, and which satisfies the classical master equation, is found. That means that the latter functional is a Hamiltonian function for the degree $-1$ symplectic structure, or else that it has vanishing BV bracket $(\cdot,\cdot)$ with itself.

The procedure developed by Batalin and Vilkovisky is highly non-unique, as it depends on several choices, so Felder and Kazhdan \cite{FelderKazhdan} suggested a list of axioms, defining a \emph{BV variety}, to prove that a solution to the classical master equation exists and the corresponding BV variety is unique up to \emph{stable equivalences} (quasi-isomorphisms).

The axioms are\footnote{The mathematical framework of ref. \cite{FelderKazhdan} is more refined than this sketch, and deals with sheaves of Poisson algebras; we will not make use of the full machinery.}:
\begin{enumerate}[label=(\roman*)]
    \item The restriction of $S$ to variables of ghost number 0 is $S_0$, the classical action in terms of physical variables;
    \item $S$ solves the classical master equation $(S,S)=0$;
    \item The cohomology in $\mA/\mI$, where $\mA$ is the full BV complex and $\mI$ is the ideal generated by variables with positive ghost number, vanishes in negative ghost number.
\end{enumerate}
They proceed to show, by a spectral sequence argument, that the axioms imply that also the cohomology in $\mA$ vanishes in negative ghost number.

The property (iii) is a natural requirement for any meaningful physical theory---non-zero cohomology except for solutions of the classical field equations and ghost zero-modes (representing global symmetries) is pathological.
Therefore it is surprising that a possible counterexample to (iii) was found in a reasonably simple model, the spinning particle with $N=1$ worldline supersymmetry \cite{Getzler:2015jrr}. Pathological cohomology at negative ghost number was explicitly constructed, which has been causing concern---is this behaviour an isolated incident, or a property of more general models involving supergravity?
The main purpose of the present paper is to show that the observed negative ghost number cohomology of the classical spinning particle is trivialised if the constraints are properly resolved in a Koszul--Tate (KT) resolution \cite{Koszul-hom,JohnTate}. Axiom (iii) of Felder and Kazhdan is then restored.

We will work in the phase space formulation of the spinning particle.
This makes the comparison with \cite{Getzler:2015jrr} straightforward.
It also gives a clear structure to the set of fields appearing in the BV action. There will be: 
\begin{itemize}
\item{One tower of ghosts (beginning with the Lagrange multipliers/gauge connections of ghost number 0) encoding the KT resolution of the constrained phase space, as a graded ring; and}
\item{One tower of ghosts encoding the $L_\infty$ algebra of transformations generated by this resolution.}
\end{itemize}
The doubling is related to the AKSZ formulation.

We work entirely in the setting of {\it classical} BV theory. 
The conclusions drawn about the resolution, and thus about the set of fields, cease to apply in the ``first-quantised'' model (see e.g.~\cite{corradini2021spinningparticlesquantummechanics}). 
``First quantisation" involves a choice of Lagrangian submanifold (polarisation) and a Hilbert space of states/wave functions there, that replaces the algebra of functions on the classical phase space. Half of the variables (and in particular of the Grassmann odd ones) become creation operators, while the other half become annihilation operators. 
Target space fields are obtained as states/wave functions in a BRST cohomology. 
This approach was taken by one of the authors in reference \cite{Boffo:2025fip}, along with setting the Lagrange multipliers to a fixed value. Simply put, there is no commutative ring structure on the constrained quantum phase space, no reducibility and no need for a (non-trivial) resolution.
It can also be seen very concretely, \eg\ in the BFV formalism \cite{GetzlerPrivate}, that quantisation trivialises the cohomology of ref. \cite{Getzler:2015jrr}.
       
In Section \ref{KTSection}, we demonstrate concretely why and how the constraints of the classical spinning particle are reducible \cite{Henneaux:1994lbw} due to the presence of fermionic variables, through ``saturation reducibility'', and give the beginning of the KT resolution. Section \ref{PartitionSec} complements and illustrates this picture with partition function arguments. Finally, in Section
\ref{KTBVSec}, the KT resolution is implemented in the BV framework for a class of 1-dimensional sigma models gauged by a superalgebra, including the spinning particle, and we identify the stage in the resolution where the cohomology of ref. \cite{Getzler:2015jrr} is trivialised. We also demonstrate and comment on the formulation of the theory as an AKSZ model.

\section{The phase space of the $N=1$ spinning particle and its resolution\label{KTSection}}

The $N=1$ spinning particle is a toy model of the RNS superstring, that sweeps a worldline rather than a worldsheet when propagating \cite{Brink:1976uf}. Its phase space is parametrised by $2d$ coordinates $x^m$, $p_m$ and $d$ Grassmann odd coordinates $\theta^m$.
The particle propagates in Minkowski space, so there is an invariant metric $\eta_{mn}$.

Let $T=p^2$, $U=p\cdot\theta$. These are the constraints of the $N=1$ spinning particle, obtained by variation of the worldline metric and gravitino field, respectively.
We are interested in the phase space modulo the constraints $T=0=U$.
The coordinates $x$ are not involved in the constraints\footnote{$x^m$ will of course transform under the local symmetries generated by the constraints; here we are just resolving the constraint surface, whereas transformations are dealt with in Section \ref{KTBVSec}.}, so we consider the (graded commutative) ring 
\begin{align}
R=\Sym^\bullet(p,\theta)/\langle T,U\rangle
\end{align}
of polynomials in $p$ and $\theta$ modulo the ideal generated by $T$ and $U$. The symmetric algebra for the parity odd $\theta$'s of course corresponds to an exterior algebra of parity even coordinates, since $\theta^m\theta^n+\theta^n\theta^m=0$ holds. 
For later purposes, it is important to resolve this quotient by enlarging the space with ghosts, in a cohomologically trivial way. Which are the appropriate ghosts? That is\footnote{This step is justified in Section \ref{KTBVSec}.}, what is the KT resolution of $R$?
(One can imagine various ways to deal with the ring $R$. It is the coordinate ring of the parity-shifted tangent sheaf of the light-c\^one.)

First, introduce ghosts $b$ (of odd parity) and $\beta$ (of even parity) for the constraints $T$ and $U$, and the differential (vector field acting on $\Sym^\bullet(p,\theta,b,\beta)$)
\begin{align}
{\mathrm d}_0=T{\*\over\*b} +U{\*\over\*\beta} \;,
\end{align}
which trivialises $T$ and $U$. 
However, the constraints are reducible.

Let $\Omega={1\over d!}\varepsilon_{m_1\ldots m_d}\theta^{m_1}\ldots\theta^{m_d}$ and
$\imath_p\Omega=(p\cdot{\*\over\*\theta})\Omega={1\over(d-1)!}\varepsilon_{m_1\ldots m_d}p^{m_1}\theta^{m_2}\ldots\theta^{m_d}=p^m\imath_m\Omega$.
It is straightforward to see that the constraints obey the reducibility 
\begin{align}
&U\Omega=0\;,\label{ReducibilityEq}\\
&T\Omega-U\imath_p\Omega=0\;,\nn
\end{align}
from antisymmetrisation in $d+1$ indices. This kind of ``saturation reducibility'' is generic for constraints involving odd variables.
The reducibility \eqref{ReducibilityEq} shows that ${\mathrm d}_0$ has cohomology represented by $\beta\Omega$ and
$b\Omega-\beta\imath_p\Omega$ (these cochains are obviously not exact). 
It is trivialised by the introduction of ghosts for ghosts $b',\beta'$ and an additional  part of the differential,
\begin{align}
{\mathrm d}_1=(b\Omega-\beta\imath_p\Omega){\*\over\*b'}+\beta\Omega{\*\over\* \beta'}\;.
\label{d1eq}
\end{align}

This is not the end. 
${\mathrm d}_0+{\mathrm d}_1$ has cohomology represented by 
\begin{align}
\omega_1&=\theta_m\beta'\;,\nn\\
\omega_2&=p_m\beta'-{1\over2}\beta^2\imath_m\Omega\;,\nn\\
\omega_3&=\theta_mb'-{1\over2}\beta^2\imath_m\Omega\;, 
\label{omega1234Eq}\\
\omega_4&=p_mb'+b\beta\imath_m\Omega+{1\over2}\beta^2\imath_m\imath_p\Omega\;,
\nn
\end{align}
which can be observed using the identities 
\begin{align}
\theta_m\Omega&=0\;,\nn\\
p_m\Omega&=U\imath_m\Omega\;,\nn\\
\theta_m\imath_p\Omega&=U\imath_m\Omega\;,\\
p_m\imath_p\Omega&=T\imath_m\Omega+U\imath_m\imath_p\Omega\;.\nn
\end{align}
These cohomologies are then trivialised by the introduction of four vector ghosts $\beta''_i$ and a term in the differential 
${\mathrm d}_2=\sum_{i=1}^4\omega_i \, \partial_{\beta''_i}$.
Part of the ensuing higher ghosts come from multiplication again with $\theta$ or $p$ and graded antisymmetrisation in the free indices.

In summary, we have been building up a few stages of the KT resolution, whose end goal is to form an acyclic complex (but in the first non-zero slot from the right)
\[
\dots \Sym^\bullet(p,\theta,b,\beta,b',\beta')\xrightarrow{{\mathrm d}_1}\Sym^\bullet(p,\theta,b,\beta) \xrightarrow{{\mathrm d}_0}\Sym^\bullet(p,\theta) \twoheadrightarrow \Sym^\bullet(p,\theta)/\langle T,U\rangle \to 0\, .
\]
Acyclicity is argued from the fact that the complex should continue ad infinitum, setting $\ker {\mathrm d}_i = \text{im} \, {\mathrm d} _{i-1}$ at every $i$.

More detailed information on the full resolution is presented in the following Section \ref{PartitionSec}, where partition functions are used to (partially) deduce the ghost spectrum of the resolution. Comparing partition functions shows that the KT resolution continues (infinitely), in powers of both the variables used. 
In fact, new powers of $\Omega$ enter at some point.
We have constructed the corresponding cohomologies/ghosts only for the first few stages. 
We are not aware of an explicit form of the full resolution, except in the somewhat degenerate case $d=1$, presented in \ref{d1Section}. All the above holds true there, but since in this case the constraints set {\it all} quadratic expression in the generators $p,\theta$ to 0, the resolution is dual to the free Lie algebra on one odd and one even generator.

\section{Partition functions\label{PartitionSec}}

\subsection{Generalities\label{PartitionGenSection}}

Besides by direct computation, information about the cohomology can be extracted also via partition functions or Hilbert--Poincar\'e series,
which compute the dimensions of cohomology groups. 
We use the terminology interchangeably, but prefer ``partition function'' when coefficients are refined to take values in some representation ring (see below).

To obtain a partition function, it is necessary to assign weights to the involved coordinates/fields in a way that the differential of the complex has weight zero. This fact ensures that the partition function of the complex and that of the cohomology are equal.
One can also consider a refined partition function (see \eg\ refs. \cite{Cederwall:2015oua,Cederwall:2023wxc}), formal power series with coefficients in the representation ring of some structure algebra, in the case of a $d$-dimensional spinning particle $\so(d)$.

The Hilbert--Poincar\'e series of an even generator of degree $n$ is $(1-t^n)^{-1}$, and that of an odd one is $1-t^n$. The series is trivially extended to more than one fugacity.
For example, consider a set of generators at degrees $n_i\in\ZZ$ with multiplicities $p_i$ (taken negative for odd variables), then the partition function of the generating set is
$F(t)=\sum_ip_it^{n_i}$, and the Hilbert--Poincar\'e series is
$P(t)=\prod_i(1-t^{n_i})^{-p_i}$.
The plethystic logarithm ($\Log$) and exponential ($\Exp$) conveniently relate partition functions and Hilbert--Poincar\'{e} series as
$P(t)=\Exp F(t)$, $F(t)=\Log P(t)$. The logarithm satisfies
$\Log(fg)=\Log f+\Log g$.

When a ring has a non-trivial degree 0 automorphism group $A$, it is convenient to replace the number $n_i$ of generators at degree $n_i$ by $A$-representations $R_i$ (this is equivalent to working with characters). We then write the refined partition\footnote{Here, we have used explicit $\oplus$ for direct sum of modules and $\otimes$ for tensor products in order to emphasise the coefficients taking values in the representation ring. Sometimes, the latter will be omitted in the following.} for the generators
$G(t)=\bigoplus_i R_it^{n_i}$ and for the ring
$Z(t)=\bigotimes_i (1-t^{n_i})^{-R_i}$. 
Here, the notation $(1-t)^{-R}$ is shorthand for
$\oplus_{k=0}^\infty \vee^k\!R\,t^k$. Odd variables are treated as even ones in $-R$, which is consistent with
$(1-t)^R=\bigoplus_{k=0}^{\dim R} \wedge^k\!R\,t^k$.
Again\footnote{This follows from the same properties of the functions of unrefined partitions by thinking of characters in terms of the exponents of fundamental weights as fugacities.}, $Z(t)=\Exp G(t)$, $G(t)=\Log Z(t)$, and
$\Log(f\otimes g)=\Log f\oplus\Log g$.

In the following, we will use the thus refined partition functions with structure group $SO(d)$. We use Young tableaux notation for the representation, with $\yng(1)$ being the vector representation. A tilde on a Young tableau (\eg\ 
$\tilde{\yng(2)}$) denotes complete tracelessness.

What happens when there are constraints on Grassmann odd variables?
Graded rings are generically less well behaved than even ones. This is due to the simple observation that multiplying enough odd variables gives 0. We may call reducibility arising for this reason a ``\emph{saturation reducibility}''.
The example below provides an illustration.

As an example,
take the simplest possible ring where some ideal is divided out. Let it be generated by two odd generators $\theta^i$, $i=1,2$, and let $T=\theta^1\theta^2$ generate the ideal. The partition function is obviously $P(t)=1-2t$; everything above degree (of homogeneity) one is in the ideal.
Now, we ask instead for the KT resolution (or Hilbert-Poincar\'{e} series) of this ring.
The generators $\theta^i$ and the (odd) ghost for the generator $T$ of the ideal contribute $(1-t)^2(1-t^2)$, which does not equal $1-2t$. Higher ghosts are needed. This is where the saturation reducibility, $\theta^iT=0$, comes into play.
The factorisation of the partition function yields an infinite product:
\begin{align}
1-2t=(1-t)^2(1-t^2)(1-t^3)^2(1-t^4)^3(1-t^5)^6\times\ldots\;.
\label{ExampleTateEq}
\end{align}
All ghosts (in this particular case) are odd. Their number matches the number of generators in a free Lie algebra $F[\CC^2]$ on 2 (even) degree one elements. 
Namely, take generators in a vector space $V$ with partition function (a polynomial without constant term) $V(t)$. The partition function of the Lie algebra $F[V]$ freely generated on $V$ (the $\Log$ of eq. \eqref{ExampleTateEq}) is complicated. However, the universal enveloping algebra of $F[V]$ is the tensor algebra $T[V]$, with partition function 
$(1-V(t))^{-1}$. 
In our example, $(1-2t)^{-1}$ is the partition function for 
$T[\CC^2]=U[F[\CC^2]]$,
the universal enveloping algebra of the free algebra over 2 elements at degree 1. This is the Lie algebra which is Koszul dual to the KT resolution, in the sense \cite{Cederwall:2015oua,Cederwall:2023wxc} that the complex of the resolution is $C^\bullet(F[\CC^2])$,
$Z(t)\otimes Z_{T[\CC^2]}(t)=1$. 
Free algebras only have cohomology $H^0$ (1) and $H^1$ (the dual generators), reflected in the left hand side of eq. \eqref{ExampleTateEq}.

In the simple example just shown, Hilbert's syzygy theorem \cite{HilbertAlgForm} does not hold. The theorem states that the additive resolution of a ring over the ring {\it freely} generated by the generators is finite.
The example above illustrates that this is not true for odd generators. The form \eqref{ExampleTateEq} of the partition function can not be written as a polynomial times $(1-t)^2$.
On the contrary, an even ring with two generators, completely constrained at quadratic level, has the Hilbert--Poincar\'e series
$1+2t$, which equals ${1-3t^2+2t^3\over(1-t)^2}$.

\subsection{The spinning particle}

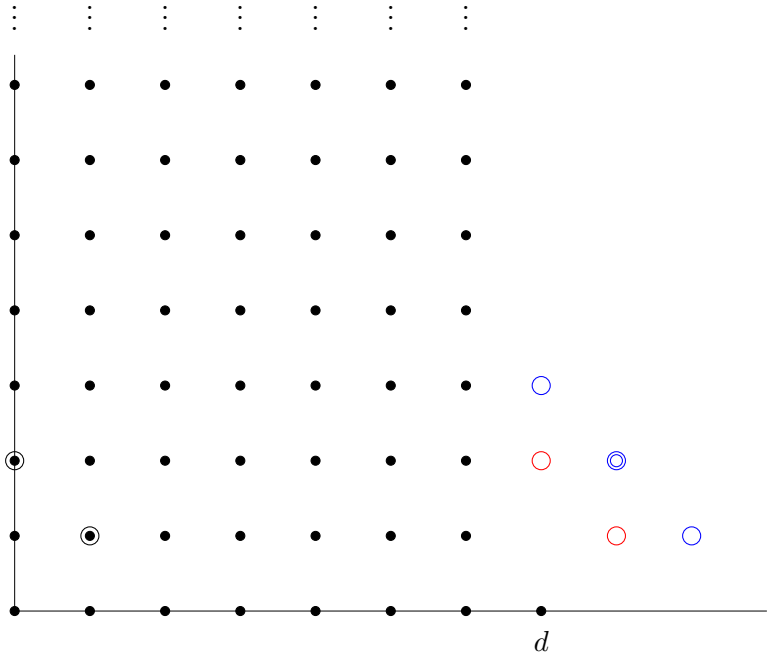
\begin{figure}[H]
\begin{center}
\begin{tikzpicture}
\draw[line width=.1pt](0,0)--(10,0);
\draw[line width=.1pt](0,0)--(0,7.4);
\node at (7,-.4) {$d$};
\foreach \a in {0,1,...,6}
\foreach \b in {0,1,...,7}
       \filldraw[black] (\a,\b) circle (0.06cm);
\filldraw[black] (7,0) circle (0.06cm);
\foreach \a in {0,1,...,6}
    \node at (\a,8){$\vdots$};
\draw(0,2) circle (0.12cm);
\draw(1,1) circle (0.12cm);
\draw(7,2)[red] circle (0.12cm);
\draw(8,1)[red] circle (0.12cm);
\draw(7,3)[blue] circle (0.12cm);
\draw(8,2)[blue] circle (0.12cm);
\draw(8,2)[blue] circle (0.08cm);
\draw(9,1)[blue] circle (0.12cm);
\end{tikzpicture}
\caption{A picture of the ring $S$. Powers of $\theta$ and $p$ are on the horizontal and vertical axis, respectively. Solid dots denote non-vanishing $\so(d)$ modules. The constraints (generators of the ideal) $T$ and $U$ are drawn as black rings, reducibility ghosts as red rings, and ghosts of the following stage as blue rings.\label{partitionFig}}
\end{center}
\end{figure}

Consider the ring $R$ for the $N=1$ spinning particle.
We define partition functions with the fugacities $s$ and $t$ associated to degree of homogeneity in $\theta$ and $p$, respectively. The partition function of $\Sym^\bullet(p,\theta)$ is
$Z_0(s,t)=(1-s)^{\expyng1}(1-t)^{-\expyng1}$
(we omit denoting obvious tensor products). We shall construct the partition function for the case  with constraints $T, U$ step-by-step.
After including ghosts for the constraints, we obtain for the first stage of the resolution,
\begin{align}\label{Z1eq}
Z_1(s,t)&=(1-s)^{\expyng1}(1-t)^{-\expyng1}(1-st)^{-1}(1-t^2)\;,
\end{align}
or, equivalently,
\begin{align}
\Log Z_1(s,t)&=-\yng(1)\,s+\yng(1)\,t+st-t^2
=(t-s)(\yng(1)-t)\label{LogZ1eq}
\end{align}
which agrees with the actual partition function of $R$ at degrees where 
the latter is non-vanishing (see Figure \ref{partitionFig}), but not at higher powers of $\theta$. 
All higher ghosts contribute with powers of $s,t$ that the partition function of $R$ does not reach. 
Restricting eq. \eqref{Z1eq} to powers $s^mt^n$, where $m\leq d$ for $n=0$ and $m\leq d-1$ for $n>0$, gives
an explicit form for the complete partition function, which however is not of the product form required for reading off the plethystic logarithm and deducing the KT resolution:
\begin{align}
Z(s,t)=(1-s)^{\expyng1}+(1-t^2)(1-t)^{-\expyng1}
\bigoplus_{m=0}^{d-1}\bigoplus_{n=1}^m(-1)^{m-n}\wedge^{m-n}\yng(1)\,s^mt^n\;.\label{FullZ}
\end{align}

Some precise statements can be made:
At $t^1$ (the next-to-lowest row in Fig. \ref{partitionFig}), we find ghosts in totally symmetric products of a vector, all with parity $d+1$. 
Let $\sigma=(-1)^d$. The partition of that row becomes $t$ times
\begin{align}
&(1-s)^{\expyng1}\otimes\left(\yng(1)\oplus s
\oplus(-\sigma) s^{d+1}(1-s)^{-\expyng1}
\right)\nn\\
&=(1-s)^{\expyng1}\otimes(\yng(1)\oplus s)\oplus(-s)^{d+1}\;,
\end{align}
which is a polynomial in $s$ of degree $d-1$. 
The leftmost column in Fig. \ref{partitionFig} is a subring, the coordinate ring of the light-c\^one. All other columns are representations of this ring, the first one consisting of (odd) tangent vectors. The whole ring can be seen as the ring of differential forms on the light-c\^one, which is the coordinate ring of the parity-shifted tangent sheaf.

The next stage involves the two ``scalar'' (supersymmetry adjoint) reducibilities. Note that the parity of these ghosts depend on whether the space-time dimension is even or odd, giving even or odd parity to $\Omega$.
At this second stage of the resolution, the reducibility ghosts that compensate for \eqref{ReducibilityEq} are taken into account, producing:
\begin{align}
\Log Z_2(s,t)&=-\yng(1)\,s+\yng(1)\,t +st-t^2
-\sigma s^{d+1}t+\sigma s^dt^2\nn\\
&=(t-s)(\yng(1)-t+\sigma s^dt)\;.
\end{align}

The following stage contains the four vectors of eq.
\eqref{omega1234Eq}. 
This can of course also be verified in the partition function. After this stage, the partition function is \begin{align}
\Log Z_3(s,t)&=-\yng(1)\,s+\yng(1)\,t +st-t^2
-\sigma s^{d+1}t+\sigma s^dt^2\nn\\
&\qquad-\sigma \yng(1)s^{d+2}t+2\sigma\yng(1)s^{d+1}t^2
-\sigma\yng(1)s^dt^3\\
&=(t-s)\left(\yng(1)-t+\sigma s^dt-\sigma\yng(1)s^dt(t-s)\right)\;.\nn
\end{align}

This is the step that we have reached with the concrete construction of the differential for the resolution in Section \ref{KTSection}.
We are not aware of a closed expression for the full resolution, except for the particular case $d=1$, see Section \ref{d1Section}.
Still, more information can be obtained by sequential factorisation of the exact partition function \eqref{FullZ}.
We have used LiE \cite{LiE} to perform this calculation up to some bidegrees for various choices of $d$.

\Yboxdim\youngexpdim
\begin{table}[h]
\renewcommand{\arraystretch}{2}
\scalebox{.65}{
\begin{tabular}{r||c|c|c|c|c|c|c|}
\cline{1-2}
$7$&$-\yng(1)-\yng(1,1,1)-\yng(1,1,1,1,1)$
\\[6pt]\cline{1-3}
$6$&$1+\yng(1,1)+\yng(1,1,1,1)$
&$\begin{matrix}3\,\yng(1)+3\,\yng(1,1,1)+2\,\yng(1,1,1,1,1)\\
    +\tilde{\yng(2,1)}+\tilde{\yng(2,1,1,1)}\end{matrix}$
\\[6pt]\cline{1-4}
$5$&$-\,\yng(1)-\,\yng(1,1,1)$
    &$\begin{matrix}-2-2\,\yng(1,1,1,1)-3\,\yng(1,1)\\
    -\tilde{\yng(2,1,1)}-\tilde{\yng(2)}\end{matrix}$
    &$\begin{matrix}-4\,\yng(1)-4\,\yng(1,1,1)-\yng(1,1,1,1,1)
    -3\,\tilde{\yng(2,1)}\\
    -\tilde{\yng(3)}-\tilde{\yng(3,1,1)}-2\,\tilde{\yng(2,1,1,1)}\end{matrix}$
    \\[6pt]\cline{1-5}
$4$&$1+\yng(1,1)$
    &$3\,\yng(1)+2\,\yng(1,1,1)+\tilde{\yng(2,1)}$
    &$\begin{matrix}2+\yng(1,1,1,1)+4\,\yng(1,1)\\
    +2\,\tilde{\yng(2,1,1)}+3\,\tilde{\yng(2)}+\yng(3,1)\end{matrix}$
    &$\begin{matrix}4\,\yng(1)+3\,\yng(1,1,1)+3\,\tilde{\yng(3)}+4\,\tilde{\yng(2,1)}\\
    +2\,\tilde{\yng(3,1,1)}+\tilde{\yng(4,1)}
    +\tilde{\yng(2,1,1,1)}\end{matrix}$
\\[6pt]\cline{1-6}
$3$&$-\,\yng(1)$&$-2-2\,\yng(1,1)-\,\tilde{\yng(2)}$
    &$\begin{matrix}-4\,\yng(1)-\yng(1,1,1)\\
    -2\,\tilde{\yng(2,1)}-\tilde{\yng(3)}\end{matrix}$
    &$\begin{matrix}-2-3\,\yng(1,1)-\tilde{\yng(2,1,1)}\\
    -4\,\tilde{\yng(2)}-2\,    \tilde{\yng(3,1)}-\tilde{\yng(4)}\end{matrix}$
    &$\begin{matrix}-4\,\yng(1)-4\,\tilde{\yng(3)}-\tilde{\yng(5)} \\-3\,\tilde{\yng(2,1)}-\yng(1,1,1)
    -\tilde{\yng(3,1,1)}-2\,\tilde{\yng(4,1)}
    \end{matrix}$
    \\[6pt]\cline{1-7}
$2$&$1$&$2\,\yng(1)$&$2+\yng(1,1)+2\,\tilde{\yng(2)}$
    &$3\,\yng(1)+\tilde{\yng(2,1)}+2\,\tilde{\yng(3)}$
    &$\begin{matrix}2+\yng(1,1)+3\,\tilde{\yng(2)}\\
    +\tilde{\yng(3,1)}+2\,\tilde{\yng(4)}\end{matrix}$
    &$\begin{matrix}3\,\yng(1)+3\,\tilde{\yng(3)}+2\,\tilde{\yng(5)}\\
    +\tilde{\yng(2,1)}+\tilde{\yng(4,1)}\end{matrix}$
\\[6pt]\cline{1-8}
$n=1$&&$-1$&$-\,\yng(1)$&$-1-\tilde{\yng(2)}$
&$-\,\yng(1)-\tilde{\yng(3)}$
&$\begin{matrix}-1-\tilde{\yng(2)}\\
    -\tilde{\yng(4)}\end{matrix}$
&$\begin{matrix}-\yng(1)-\tilde{\yng(3)}\\
    -\tilde{\yng(5)}
\end{matrix}$
\\[6pt]\hline\hline
&$m=d$&$d+1$&$d+2$&$d+3$&$d+4$&$d+5$&$d+6$
\end{tabular}
}
\caption{Coefficients of $(-1)^ds^mt^n$ in $\Log Z(s,t)$ for stages 2 to 7 of the KT resolution. These modules are always present, but further ghosts appear at $m\geq2d$, $n\geq2$. The ghosts at stage $2+j$,
$j\geq0$, appear at $m+n=d+j$, and are in modules contained in 
the $j$-fold tensor product of a vector
with some multiplicities, of course forming 
some modules of the 1-dimensional super-translation algebra.
Tilde means traceless. Note the ``duality'' under reflection in the diagonal $m-n=d-1$ and reflection of {\it tilded} Young tableaux.
\label{ghosttable}}
\end{table}
\Yboxdim\youngdim

One subset of ghosts, listed in Table \ref{ghosttable}, that may be identified is obtained as the tensor product of the module of the first reducibility ghosts (partition function $\sigma(t-s)s^dt$) with 
graded antisymmetric tensor products of the coordinate module (partition function $(t-s)\yng(1)$).
New cohomologies arise due to graded antisymmetrisation in $\xi_I=(p,\theta)$. By induction, suppose there are ghosts $b_{[I_1\ldots I_k)}$, and that the differential contains
$\xi_{I_1}b_{I_2\ldots I_k}{\*\over\* b_{I_1\ldots I_k}}$.
Then, the next cohomology to eliminate is
$\xi_{[I_1}b_{I_2\ldots I_{k+1})}+\ldots$, since 
$\xi_{[I}\xi_{J)}=0$. 
This alone  would give a partition of the reducibility ghosts
\begin{align}
&(-1)^d(s^dt^2-s^{d+1}t)\bigoplus_{k=0}^\infty
\vee^k(\yng(1)\,s-\,\yng(1)\,t)\nn\\
&=(-1)^ds^dt(t-s)(1-s)^{-\xyng}(1-t)^{\xyng}\;.\label{gradasEq}
\end{align}
One obtains lists of Young tableaux of a simple hook structure (no \Yboxdim\youngexpdim$\yng(2,2)\,$\Yboxdim\youngdim sub-tableaux).
However, this does not account for the smaller modules appearing together with the antisymmetric one, \eg\ in the first column of Table \ref{ghosttable}. They are due to traces also being allowed when multiplying with $p$, leading to $T$.      
The result is an infinite number of copies of eq. \eqref{gradasEq}, shifted an even number of steps upwards:
\begin{align}
(-1)^d{s^dt(t-s)\over1-t^2}(1-s)^{-\xyng}(1-t)^{\xyng}\;.
\end{align}
This leads to the contribution to the logarithm of the partition function
\begin{align}
\Log Z^{(1)}(s,t)=-\,\yng(1)\,s+\yng(1)\,t+st-t^2
+(-1)^d{s^dt(t-s)\over1-t^2}(1-s)^{-\xyng}(1-t)^{\xyng}+\ldots\;.
\label{LogZ1Eq}
\end{align}
It is interesting that the reducibility part already looks like a Hilbert space: terms such as $(1-s)^{-\xyng}$ look like a partition function of something simple (not $\Log$ of it). 
We have checked in a number of examples (specific values of $d$) that the plethystic exponentiation of eq. \eqref{LogZ1Eq},
\begin{align}
Z^{(1)}(s,t)=\left({1-s\over1-t}\right)^{\xyng}{1-t^2\over1-st}
\bigotimes_{i,j,k=0}^\infty
\left({1-s^{d+i+1}t^{j+2k+1}\over 1-s^{d+i}t^{j+2k+2}}\right)^{(-1)^{d+j}(\vee^i\xyng)\otimes(\wedge^j\xyng)}
\;,\label{Z1Eq}
\end{align}
has support on the black dots of Figure \ref{partitionFig} and on $m\geq 2d$, $n\geq2$.

Comparison of eq. \eqref{Z1Eq} with the exact partition function (in some examples) reveals that yet more ghosts appear at $m\geq2d$, which indicates further appearances of $\Omega$. The first few contributions to the series of these ghosts differ depending on the dimension $d$ being even or odd, and are depicted in Table
\ref{ghosttable2}. These are again followed by diagonals containing modules in the tensor product of increasing numbers of vectors.
It is likely that the full structure carries a filtration on the number of $\Omega$'s, the ``sheet'', needed to sequentially reach a ghost, so that the coordinates are in sheet 0, the ghosts of Table \ref{ghosttable} in sheet 1 (the superscript in eq. \eqref{LogZ1Eq} refers to this number), and 
the ones in Table \ref{ghosttable2} in sheet 2, etc.

\begin{table}[h]
\renewcommand{\arraystretch}{2}
\scalebox{.8}{
\begin{tabular}{r||c|c|c|c|}
\cline{1-2}
$5$&$\yng(1)$
\\[6pt]\cline{1-3}
$4$&&$-3\,\yng(1)$\\[6pt]\cline{1-4}
$3$&&$1$&$3\,\yng(1)$\\[6pt]\cline{1-5}
$n=2$&&&$-1$&$-\,\yng(1)$
\\[6pt]\hline\hline
&$m=2d$&$2d+1$&$2d+2$&$2d+3$
\end{tabular}
}
\hspace{36pt}
\scalebox{.8}{
\begin{tabular}{r||c|c|c|c|}
\cline{1-2}
$5$&$\yng(1)$
\\[6pt]\cline{1-3}
$4$&$-1$&$-3\,\yng(1)$\\[6pt]\cline{1-4}
$3$&&$1$&$3\,\yng(1)$\\[6pt]\cline{1-5}
$n=2$&&&&$-\,\yng(1)$
\\[6pt]\hline\hline
&$m=2d$&$2d+1$&$2d+2$&$2d+3$
\end{tabular}
}
\caption{The beginning of further ghosts (beyond those in Table \ref{ghosttable}), due to a second appearance of $\Omega$. The left table is for odd $d$, the right one for even $d$.
\label{ghosttable2}}
\end{table}

Although we are not aware of a closed multiplicative form for the partition function, in Section \ref{ResSec} we will present the classical BV formulation of a physical theory with reducible constraints (such as a particle with world-line supersymmetry) and in Section \ref{CompSS} we will demonstrate explicitly how the cohomologies observed in ref. \cite{Getzler:2015jrr}, which violate Felder--Kazhdan, are trivialised by the reducibility ghosts. Before that, we wish to discuss the simple, 1-dimensional situation, where the resolution can be computed explicitly at all stages.

\subsubsection{Reducibility of the 1-dimensional model\label{d1Section}}

The only case when we know the complete resolution is when $d=1$. Then $Z(s,t)=1-s+t$. 
The ring is completely constrained at quadratic level, and the resolution is the complex $C^\bullet(\fh)$, where $\fh$ is the freely generated Lie superalgebra on one even and one odd generator
(by the same argument as in the example in Section
\ref{PartitionGenSection}).
Even then, the list of generators in $\Log Z$ becomes quite intricate, see Table \ref{d1resolution}. 
It is intriguing to note that the numbers on the diagonals at
$m+n=1,3,4,5,7,8,9,11,\ldots$ are reflection symmetric around $m=n$, but the ones at $m+n=2,6,10,\ldots$ are not.
$d=1$ is the only case when the ring is Koszul, \ie, when the resolution is dual to (the universal enveloping algebra of) a Lie superalgebra. For higher $d$, the $L_\infty$ algebra ``Koszul dual'' to the resolution, 
(in the sense that the cochain complex of the $L_\infty$ algebra is the complex of the resolution)
has at least a $(d+1)$-bracket, as seen in eq. \eqref{d1eq}. 

\begin{table}[H]
\renewcommand{\arraystretch}{2}
\begin{tabular}{r||c|c|c|c|c|c|c|c|c|c|c|}
\cline{1-3}
$10$&&$-1$\\[4pt]
\cline{1-4}
$9$&&1&5\\[4pt]
\cline{1-5}
$8$&&$-1$&$-4$&$-15$\\[4pt]
\cline{1-6}
$7$&&1&$4$&$12$&$30$\\[4pt]
\cline{1-7}
$6$&&$-1$&$-4$&$-9$&$-22$&$-42$\\[4pt]
\cline{1-8}
$5$&&1&3&7&14&25&42\\[4pt]
\cline{1-9}
$4$&&$-1$&$-2$&$-5$&$-8$&$-14$&$-20$&$-30$\\[4pt]
\cline{1-10}
$3$&&1&2&3&5&7&9&12&15\\[4pt]
\cline{1-11}
$2$&$-1$&$-1$&$-2$&$-2$&$-3$&$-3$&$-4$&$-4$&$-5$&$-5$\\[4pt]
\cline{1-12}
$1$&$1$&1&1&1&1&1&1&1&1&1&$\,1\,$\\[4pt]
\cline{1-12}
$n=0$&&$-1$&&&&&&&&&
\\[4pt]\hline\hline
$m=$&$0$&$1$&$2$&$3$&$4$&$5$&$6$&$7$&$8$&$9$&$10$
\end{tabular}
\caption{The beginning of the complete resolution (dual to a free Lie superalgebra) for $d=1$.
\label{d1resolution}}
\end{table}

\vspace{1cm}

\section{Koszul--Tate resolution and BV\label{KTBVSec}}

\subsection{The resolution\label{ResSec}}

After having completed an extensive classical analysis in phase space, in this part we 
address the 
BV formulation, show the equivalence between KT and the BV-BRST complex 
and discuss how axiom (iii) of Felder--Kazhdan is fulfilled by the spinning particle example.

Our starting point will be a generic mechanical phase space/first order formalism action $S_{(0)}$ with graded phase space coordinates 
$x^i$, constraints $T_\alpha(x)$ and Lagrange multipliers $A^\alpha$,
\begin{align}
S_{(0)}=\tint {1\over2}x^i\dot x^j\omega_{ji}-H(x)+ A^\alpha T_\alpha(x)\;.
\end{align}
Applied to the spinning particle, our variables $x^i$ label the phase space variables $(x,p,\theta)$ of Section \ref{KTSection}, while $A$ refers to the Lagrange multipliers $e,\psi$ (the vielbein and its superpartner).
All variables depend on the world-line coordinate    $t$, and integration is denoted $\int=\int dt$, just to remember when total derivatives may be dropped.

Note that ordering is important when objects are graded. Our conventions are explained in detail in Appendix \ref{GradedApp}; most importantly, all index contractions are performed NW-SE and with nearby indices. 
Note that we are treating the full target phase space as a graded variety with the action of a superalgebra, while the worldline is not extended with odd variables. We are thus not using a worldline superfield formalism.
We work with $\omega_{ij}$ in a Darboux chart. The non-degenerate symplectic form $\omega$ is graded antisymmetric, 
$\omega_{ij}=(-1)^{ij}\omega_{ji}=(-1)^i\omega_{ji}$.  
In the spinning particle case, $H=0$, which is used in the following.

For quadratic constraints in the coordinates, which we assume, it is convenient to think of the Lagrange multiplier term as a minimal coupling to a gauge connection $A^\alpha$ valued in a superalgebra\footnote{For the spinning particle model at hand, $\fg$ is the $N=1$ supertranslation algebra in one dimension.} $\fg$, so that
\begin{align}
S_{(0)}=\tint {1\over2}x^iDx^j\omega_{ji}\;,
\end{align}
where $Dx^i=\dot x^i+x^j A^\alpha t_{\alpha j}{}^i$.
Consequently (see Appendix \ref{GradedApp}), fields in arbitrary $\fg$ modules have
$D\phi=\dot\phi-A\cdot\phi$, in particular for the coadjoint $D\phi_\alpha=\dot\phi_\alpha-A^\beta t_{\beta\alpha}{}^\gamma\phi_\gamma=\dot\phi_\alpha-A^\beta f_{\alpha\beta}{}^\gamma\phi_\gamma$.
The model is a gauged 1-dimensional $\sigma$-model.  
By moving $A$ to the left, we extract
\begin{align}
T_\alpha={1\over2}(-1)^\alpha x^ix^jt_{\alpha i}{}^k\omega_{kj}\;.
\end{align}

The action functional $S_{(0)}$ is degenerate because of gauge symmetries. To remove this degeneracy, we can crank the same machinery as in sec. \ref{KTSection}, with the difference that now instead of ghosts and their differentials we deploy BV conjugate pairs of coordinates\footnote{Due to the $(-1)$-shifted symplectic structure of BV, the degrees---both ghost number and parity---of a coordinate $y$ and its conjugated $y^\star$ obey $\vert y \vert + \vert y^\star \vert = -1$.}. After the ghosts $c^\alpha$ and $c^\star_\alpha$ are introduced, the BV action is $S_{(0)}+S_{(1)}+S_{(2)}$, where
\begin{align}
S_{(1)}&=\tint c^\alpha\left((-1)^{i\alpha}x^it_{\alpha i}{}^jx^\star_j+DA^\star_\alpha\right)
\equiv \tint c^\alpha\Phi_\alpha\;,
\nn\\
S_{(2)}&= \tint -{\frac 1 2}(-1)^\alpha c^\alpha c^\beta t_{\beta\alpha}{}^\gamma  c^\star_\gamma\;.\label{S1S2Eq}
\end{align}

For the odd BV bracket we take the convention\footnote{$F,G$ are any functions of the phase space coordinates enriched by the ghosts and their conjugates. $\chi^I$ collectively labels all coordinates and ghosts, so for the moment $\chi = (x, A,c)$, but the space will be extended by the resolution.}:
\begin{align}
(F,G) = F\left({\tint} \Cev{\delta\over\delta \chi^\star_I}\Vec{\delta\over\delta \chi^I}-\Cev{\delta\over\delta \chi^I} \Vec{\delta\over\delta \chi^\star_I}\right)G\; .
\end{align}
Note that the odd symplectic form could be taken as either a superform or an integral form--a distributionally valued form in supergeometry \cite{BerLei77,Belopolsky:1997bg}. We will stick with the former in our investigation.
The contraction in the second term is opposite to our superalgebra ordering conventions, but this fact does not give a sign error since $\chi^I$ and $\chi^\star_I$ have opposite parity.
Then, using $F\cev{{\*\over\*x}}=(-1)^{|x|(|F|+1)}\vec{\*\over\*x}F$, one can appreciate the graded antisymmetry of the bracket:
\begin{align}
(F,G)=-(-1)^{(|F|+1)(|G|+1)}(G,F)\;.
\end{align}

It is then straightforward, although a bit tedious due to signs, to show that $S= S_{(0)} + S_{(1)} + S_{(2)} $ has null BV bracket with itself.
The main observation is that $\Phi_\alpha$ represents cohomology of $d_{(0)}=(\cdot,S_{(0)})$, resolved on $\mA/\mI$ by the introduction of $c,c^\star$ and $S^{(1)}$ and that they generate $\fg$:
\begin{align}
(\tint c^\alpha\Phi_\alpha,\tint c^\beta\Phi_\beta)
=\tint (-1)^\alpha c^\alpha c^\beta t_{\beta\alpha}{}^\gamma \Phi_\gamma\;.
\end{align}

Had the constraints/gauge transformations been irreducible, $S_{(0)} + S_{(1)} + S_{(2)}$ would have been the complete BV action. Instead, we know that they are reducible, which should be properly taken into account. We now set out to demonstrate:

(1) that the elimination of negative ghost number cohomology in $\mA/\mI$ by the introduction of ghosts (and their antifields) in the BV framework agrees with the KT resolution of Section \ref{KTSection};

(2) that the negative ghost number cohomology encountered in ref. \cite{Getzler:2015jrr} derives from omission of higher ghosts (reducibility) and is trivialised by the resolution.

We will rely on the theorem of ref.
\cite{FelderKazhdan} that absence of negative ghost number cohomology in $\mA/\mI$ implies the same in $\mA$.
The terms $\mA$ and $\mA/\mI$ will be used not only for the final (full) complex, but also for the complexes obtained after some stage in the resolution.

The procedure \cite{FelderKazhdan} is to construct the resolution by compensating for cohomology of negative ghost number in the ring 
$\mA/\mI$, where $\mA$ consists of polynomials (or similar) in all BV coordinates and $\mI$ is the ideal generated by those of positive ghost number, so that 
the result is vanishing cohomology $H^\bullet(\mA/\mI)$ in non-zero (\ie, negative) ghost number\footnote{The differential on $\mA$ is $(\cdot,S)$; the differential on $\mA/\mI$ is $(\cdot,S)$ mod $\mI$.}. 
It is shown in ref. \cite{FelderKazhdan} that this implies also that $H^\bullet(\mA)$ vanishes in negative ghost number.
Passing to the ideal makes the (sequential) resolution possible: without taking the ideal, terms that appear at a later stage in a KT resolution may play a r\^ole at an earlier one if ghost number is unbounded in both directions. If we are able to do so, then the axioms \cite{FelderKazhdan} hold and therefore $S_0$ admits a unique (up to quasi-isomorphism) BV extension $S$, $(S,S)=0$.

Suppose that there are relations 
\begin{align}
h_{\alpha'}{}^\alpha(x) T_\alpha(x)=0\label{hReducibilityEq}
\end{align}
 in some $\fg$-module labelled by the index $\alpha'$.
Let us investigate the consequences of eq. \eqref{hReducibilityEq}.
Naively, one expects expressions $h_{\alpha'}{}^\alpha A^\star_\alpha$ to represent cohomology. Let us calculate
\begin{align}
(h_{\alpha'}{}^\alpha A^\star_\alpha,S)=h_{\alpha'}{}^\alpha(T_\alpha+(-1))^\alpha c^\beta t_{\beta\alpha}{}^\gamma  A^\star_\gamma)
+(-1)^i x^i c^\alpha t_{\alpha i}{}^j\*_jh_{\alpha'}{}^\beta A^\star_\beta\;.
\label{coho}
\end{align}
The first term vanishes, and the terms with $c A^\star$ vanish in $\mA/\mI$.  

So one needs to introduce reducibility ghosts $ A'^{\alpha'}$ and their antifields $ A'^\star_{\alpha'}$ of ghost numbers 1 and $-2$, respectively, to trivialise the cohomology $[h_{\alpha'}{}^\alpha  A^\star_\alpha]$. 
To be explicit, the new BV differential $(\cdot,S')$ is made up with
\begin{align}
S'= 
S+\tint A'^{\alpha'} h_{\alpha'}{}^\alpha A^\star_\alpha +(-1)^{\alpha'}
A'^{\alpha'}c^\alpha t_{\alpha\alpha'}{}^{\beta'} A'^\star_{\beta'}\;.
\label{SprimeEq}
\end{align}

Note that only $S_{(0)}$ was used to find the reducibility killed by the introduction of $ A'$ (since the other terms in the differential on $x$ and $ A^\star$ belong to $\mI$, see eq. \eqref{coho}). 
We want to compare the procedure here to the KT resolution of Section \ref{KTSection}. $ A^\star$ carries ghost number $-1$, just like the anti-ghosts $b,\beta$ in Section \ref{KTSection}. The trivialisation of the cohomology (on $\mA/\mI$) $h_{\alpha'}{}^\alpha A^\star_\alpha$ by the introduction of $ A'^\star_{\alpha'}$ (ghost number $-2$) precisely corresponds to the trivialisation of cohomology by introducing $b',\beta'$ (ghost number $-2$) in eq. 
\eqref{d1eq}. Similarly, by sequentially eliminating remaining cohomology on $\mA/\mI$ we observe that we get a sequence of anti-fields $ A^\star, A'^\star, A''^\star,\ldots$ with ghost numbers $-1,-2,-3,\ldots$, precisely corresponding to higher anti-ghosts in the KT resolution. We are in fact performing the KT resolution of the algebraic constraint surface in phase space, in the BV framework.

Above, we have focused on the sequence of ghosts and ghosts antifields responsible for the resolution of the constraint surface.
There will of course be a parallell tower of ghosts for the transformations, \ie, accompanying $c_\alpha$.
Next is the introduction of ghosts $c'^{\alpha'}$.
When $ A'^{\alpha'}$ is introduced, a new cohomology arises.
The differential $\mathrm{d}' =(\cdot,S')$ on $\mA/\mI$ is, so far, 
\begin{align}
\mathrm{d}'x^\star_i&=
Dx^j\omega_{ji}\;,\nn\\
\mathrm{d}' A^\star_\alpha&= 
T_\alpha\;,\label{SmallSEq}\\
\mathrm{d}' A'^\star_{\alpha'}&=h_{\alpha'}{}^\alpha A^\star_\alpha\;,\nn\\
\mathrm{d}'c^\star_\alpha&=(-1)^{i\alpha}x^i t_{\alpha i}{}^jix^\star_j
    +D A^\star_\alpha\nn \; .
\end{align}
A short calculation now finds a cohomology at ghost number $-2$:
\begin{align}
\Omega_{\alpha'}=h_{\alpha'}{}^\alpha c^\star_\alpha-D A'^\star_{\alpha'}
-(-1)^{i+\alpha'}\tilde\omega^{ij}x^\star_j\*_ih_{\alpha'}{}^\alpha A^\star_\alpha\;.
\label{SInvEq}
\end{align}
The last term is due to the transformation of $h_{\alpha'}{}^\alpha$. $\tilde\omega$ is the super-inverse of $\omega$,
$\tilde\omega^{ik}\omega_{kj}=(-1)^i\delta^i{}_j$.
This is straightforwardly checked using eq. \eqref{SmallSEq}. One needs the $\fg$-transformation of $h_{\alpha'}{}^\alpha$, encoded in
$Dh_{\alpha'}{}^\alpha=Dx^i\*_ih_{\alpha'}{}^\alpha$, 
as well as the identity
\begin{align}
0=\*_i(h_{\alpha'}{}^\alpha T_\alpha)
=\*_ih_{\alpha'}{}^\alpha T_\alpha
+(-1)^{\alpha+i(\alpha'+\alpha)}h_{\alpha'}{}^\alpha x^kt_{\alpha k}{}^j\omega_{ji}\;.
\end{align}
The cohomology $\Omega_{\alpha'}$ is then killed by the introduction of $c'^{\alpha'}$, $c'^\star_{\alpha'}$, and 
$S''=S'+\tint c'^{\alpha'}\Omega_{\alpha'}+\ldots$, 
$(c'^\star_{\alpha'},S'')=\Omega^{-2}_{\alpha'}+\ldots$, the ellipsis denoting terms in $\mI$.
This of course has to be lifted to $\mA$, introducing the beginning of the brackets of the (graded) $L_\infty$ algebra generated by the resolved constraints.

\subsection{Comparison with ref. \cite{Getzler:2015jrr}\label{CompSS}}

When we specialise to the spinning particle, the Lie superalgebra $\fg$ is the $N=1$ supertranslation algebra in 1 dimension. The only non-vanishing bracket is 
$[U,U]=T$. The adjoint module, as well as the coadjoint, is spanned by one odd and one even generator, of which one transforms into the other under $U$, and both map to 0 under $T$. Therefore, the coadjoint is isomorphic to the parity-shifted adjoint, $\mathfrak{g}^* = \mathfrak{g}[1]$. This is also seen from the invariance of the bilinear form $T\otimes U-U\otimes T$.

 The pair $(\Omega,\imath_p\Omega)$ is also in the adjoint of $\fg$ (with parity depending on whether $d$ is even or odd, since $\Omega$ is even for even $d$ and odd for odd $d$). It is straightforward to see that the tensor product of the (co-)adjoint with itself contains the adjoint, not only using the structure constants, but also using another invariant tensor leading to the expressions
 $\beta\Omega$ and $b\Omega-\beta\imath_p\Omega$ of Section \ref{KTSection}. In the BV framework, the ghosts $b,\beta$ are replaced by the antifields of the Lagrange multipliers $A^\alpha$. For the spinning particle, they are then the antifields $e^\star,\psi^\star$ of the world-line einbein $e$ and its superpartner $\psi$.
We thus have the (degree $-1$) cohomologies in $\mA/\mI$ to be killed represented by $\psi^\star\Omega$ and 
$e^\star\Omega-\psi^\star\imath_p\Omega$.
The latter one agrees with a cohomology found in ref. \cite{Getzler:2015jrr} as ``$\zeta_1(1)$'' on p.11, while the first one is not present there. Let us go into some detail why this happens.

The cohomology observed in ref. \cite{Getzler:2015jrr}
is constructed as the differential (in our setting $(\cdot,S)$) acting on some singular function on the BV variety (more precisely, containing the inverse of the super-translation ghost $\gamma$), but in such a way that the resulting function is regular\footnote{This is independent of the prescription for such functions: $C^\infty$, polynomial, formal power series.}.
A set of cohomologies found in ref. \cite{Getzler:2015jrr}, there represented by $\zeta_k(f)$, take the form, when restricted to $\mA/\mI$ and adapted to our conventions,
\begin{align}
\zeta_k(f)=k(\psi^\star)^{k-1}e^\star f\Omega-(\psi^\star)^kf\imath_p\Omega\;.
\end{align}
Here, $1\leq k\in\ZZ$ and $f$ is a function of the coordinates\footnote{Not to confuse with the phase space coordinates $x^i$ of Section \ref{ResSec}.} $x^m$, which becomes a mere spectator on $\mA/\mI$.
Setting $k=1$, $f=1$ gives 
$\zeta_1(1)=e^\star\Omega-\psi^\star\imath_p\Omega$, a cohomology trivialised by the introduction of reducibility ghosts.
Then clearly, any of the cohomologies $\zeta_k(f)$ can be written as a linear combination of $\psi^\star\Omega$ and 
$e^\star\Omega-\psi^\star\imath_p\Omega$, and will be trivialised by the introduction of the reducibility ghosts. Thus they cannot represent cohomology when the resolution of Section \ref{KTSection}
is performed---quite trivially, since that resolution removes all cohomology in $\mA/\mI$ at negative ghost number\footnote{An explicit construction of the function $\xi_k(f)$ such that $\zeta_k(f)=d\xi_k(f)$ will involve higher reducibility ghosts. For example, $\zeta_2(1)=\psi^\star de'^\star+e^\star d\psi^\star\simeq Ue'^\star-T\psi'^\star$. Forming $\xi_2(1)$ involves the four vectors in the following stage of the resolution.}.

The cohomology $\psi^\star\Omega$ is not observed in ref. \cite{Getzler:2015jrr}.
The reason is the following: A cohomology encoding reducibility on $\mA/\mI$ persists in $\mA$ (without the introduction of new terms in the differential) if it is a $\fg$-scalar, given the construction in Section \ref{ResSec}. Then $t_{\alpha\alpha'}{}^{\beta'}=0$ for the corresponding value of $\beta'$, and the last term in $S'$, eq. \eqref{SprimeEq}, does not contribute to $\mathrm{d}'A'^\star_{\beta'}$ which remains in $\mA/\mI$. This holds for $e^\star\Omega-\psi^\star\imath_p\Omega$, being in the ``$T$'' part of the adjoint of the supersymmetry algebra, but not for 
$\psi^\star\Omega$, which needs to wait for the introduction of the reducibility ghosts. Since these are not present in ref. \cite{Getzler:2015jrr}, the cohomology does not persist.
 
Note that the cohomology killed by $c',c'^\star$ is not observed in ref. \cite{Getzler:2015jrr}, since they already require the introduction of $ A', A'^\star$.

Ref. \cite{Getzler:2015jrr} also finds cohomologies denoted by $\alpha_k(f)$ with $k>1$.
They are constructed in a way analogous to $\zeta_k(f)$.
However, they lie in the ideal $\mI$. For example,
\begin{align}
\alpha_k(1)=(\psi^\star)^k\bigl(\gamma\psi^\star\Omega
+c(e^\star\Omega-\psi^\star\imath_p\Omega)\bigr)\;.
\end{align}
Linear combinations of $\psi^\star\Omega$ and 
$e^\star\Omega-\psi^\star\imath_p\Omega$ again appear.
Since they belong to $\mI$, ref. \cite{FelderKazhdan} guarantees that we need not eliminate them by the introduction of new ghosts, they will trivialise when the differential is extended to $\mA$.

\subsection{AKSZ formulation\label{AKSZSection}}

The Alexandrov--Kontsevich--Schwarz--Zaboronsky (AKSZ) construction \cite{Alexandrov:1995kv} is a celebrated algorithm that allows to obtain a BV theory on the maps between a dg manifold and a QP manifold (dg with a compatible homological vector field).

The target QP manifold $\mathscr N$ in the present setting of an $N=1$ spinning particle, has coordinates $X^A$ at the first line in Table \ref{AKSZFieldsTable}.
The source dg manifold $\Sigma$ is $T[1]\RR$, with coordinates $(t,\eta)$ and differential $\db=\eta{d\over dt}$.
It clear from the list of coordinates that 
there is a degree 0 symplectic structure $\Omega$ 
on $\NC$,
extending the one of phase space with coordinates $x^i$.

\renewcommand{\arraystretch}{1.4}
\begin{table}[h]
\begin{center}
\begin{tabular}{r|cccccccccc}
gh\# &$\ldots$&$-4$&$-3$&$-2$&$-1$&0&1&2&3&$\ldots$\\ \hline
$X^A$&&$\ldots$&
$A''^\star_{\alpha''}$&$A'^\star_{\alpha'}$&$A^\star_\alpha$&$x^i$&$c^\alpha$&$c'^{\alpha'}$&$c''^{\alpha''}$&$\ldots$\\
$X^\star_A$&$\ldots$&$c''^\star_{\alpha''}$
&$c'^\star_{\alpha'}$&$c^\star_\alpha$&$x^\star_i$&$A^\alpha$&$A'^{\alpha'}$&$A''^{\alpha''}$&$\ldots$
\end{tabular}
\caption{The BV fields of the model, organised to display the symplectic structure. \label{AKSZFieldsTable}}
\end{center}
\end{table}
\renewcommand{\arraystretch}{1}

A map $Y:\Sigma\rightarrow\NC$ is given by
\begin{align}
Y^A(t,\eta)=X^A(t)+\eta\,\tilde\Omega^{AB}X^\star_B(t)\;,
\end{align}
with $\tilde\Omega$ being the super-inverse of $\Omega$ (used for example in eq. \eqref{SInvEq}). The ghost number is the sum of degree of $\eta$ and the degree of the element of the QP manifold. $\tilde\Omega^{AB}X^\star_B$ provide coordinates on $\NC[-1]$ (see Table \ref{AKSZFieldsTable}).

The AKSZ construction uses the differentials on the target as well as on the source. Let the former be $q$, generated from a BRST charge\footnote{The BRST charge naturally combines with the Hamiltonian in a superfield $Q+\eta H$. In our setting, $H=0$.} $Q$ as $q=\{\cdot,Q\}$ by the Poisson bracket on $\NC$. 
The BV action is constructed as 
\begin{align}
S=\int_{T[1]\mathbb{R}} dt\,d\eta\,\left({1\over2}Y^A\db Y^B\Omega_{BA}+Q\right)\;.
\label{AKSZAction}
\end{align}
which then produces the BV differential as the sum of the source differential and the pullback of the target one. Hamiltonian mechanics as AKSZ, however without reducibility, has been considered in
refs. \cite{Grigoriev_2000,Basile:2025ylc}.

We want to verify that the system at hand has an AKSZ formulation. The kinetic terms (containing time derivatives) in Section \ref{ResSec} are directly seen to reflect the symplectic structure on $\NC$, and are obtained as the first term in eq. \eqref{AKSZAction}. In order to verify that all other terms in our BV action correspond to a BRST operator on $\NC$,
the only necessary check is that $(\cdot,S)$ maps $C^\infty (\NC)\rightarrow C^\infty(\NC)$. This is equivalent to all non-kinetic terms in $S$ being linear in $X^\star_A$ (the coordinates on the second line in Table \ref{AKSZFieldsTable}).
This is indeed the case for the terms we have written.
It is also straightforward to realise that this property will persist to all orders of reducibility, from the way new ghosts are introduced to kill cohomology. Remember that the coordinates $A^\star$ (at some level of reducibility) replace the anti-ghosts $b$ of \cref{KTSection} and all terms in the differential of the resolution are linear in derivatives ${\*\over\*b}$.
For the coordinates $c$ (at some level of reducibility), they are coalgebra 1-forms in the cochain complex of an $L_\infty$ algebra (also acting on the other coordinates). Both these statements imply linearity in $X^\star$.

Finally, it was shown in ref. \cite{Barnich:2009jy} that a kind of Poincar\'e lemma holds, so the cohomology of $s=(\cdot,S)$ on the maps $\Sigma\rightarrow\NC$ is the cohomology of the differential $q$ on $\NC$. This is equivalent to say that there is a quasi-isomorphism between  $C^\infty(\NC)$ and functionals over $\text{Maps}(\Sigma, \NC)$.
Hence in the present case, after finding the appropriate resolution of the constraints, we could have alternatively used Barnich--Grigoriev's theorem to infer that there is cohomology only in the physical ghost number 0 phase space (apart from possible ghost zero-modes).
The extension to $\NC$ of the algebraic resolution to include the ghosts $c$ at positive ghost numbers is appealing. In the sequence
\begin{align}
\ldots\rightarrow A'^\star\rightarrow A^\star \rightarrow x\rightarrow c\rightarrow c'\rightarrow\ldots
\end{align}
(which should not be seen as linear maps, but maps to functions of the corresponding degree, polynomial in any coordinate of non-zero degree)
while $A^\star\rightarrow x$ implements the algebraic constraints, $x\rightarrow c$ takes care of the gauge symmetry, and outer arrows account for reducibility, both of the algebraic constraints and of the transformations.

\section{Conclusions}

The axioms in ref. \cite{FelderKazhdan} are natural and should apply to any meaningful physical model, including the $N=1$ spinning particle. 
We have demonstrated how the pathological cohomology observed in ref. \cite{Getzler:2015jrr} is trivialised by a proper resolution of the classical phase space, and revealed some aspects of this resolution.

We do not expect the results to have any direct applications---the main uses of particle actions are in the quantum setting, providing free fields (``first quantisation'').
However, we note that any classical model with constraints involving odd (fermionic) variables will exhibit analogous behaviour, involving saturation reducibility.
This includes, for example, RNS strings, superstrings and classical supergravities. Caution must be observed when calculating classical cohomologies in such models.

\subsection*{Acknowledgements}
First and foremost, we are indebted to Ezra Getzler for his bright insights and inputs. We are also grateful to Chris Hull and Ivo Sachs for informative discussions.
Part of the research presented was conducted during the program ``Cohomological Aspects of Quantum Field Theory'' at the Mittag-Leffler Institute, Djursholm, Sweden, supported by the Swedish Research Council under grant no. 2021-06594. E.B.~is supported by GA\v{C}R grant PIF-OUT 24-10634O.

\appendix

\section{Tensor conventions for superalgebras and their modules\label{GradedApp}}

A Lie superalgebra $\fg$ has generators of even and odd parity ($\ZZ_2$ grading). We use a notation where for brevity, in sign factors, the parity (0 for even, 1 for odd) of a generator $T_\alpha\in\fg$ is denoted by the index $\alpha$ itself\footnote{Other common conventions is to write $p(\alpha)$ or $|\alpha|$. Since the only place they appear are in exponents of $-1$, there should be no risk of ambiguity.}. The Lie bracket $[\cdot,\cdot]$ is graded antisymmetric, \ie, $[T_\alpha,T_\beta]=-(-1)^{\alpha\beta}[T_\beta,T_\alpha]$. The Jacobi identities read
$[[T_{[\alpha},T_\beta],T_{\gamma]}]=0$, $[\alpha\beta\gamma]$ denoting total graded antisymmetrisation. Structure constants are defined by 
$[T_\alpha,T_\beta]=f_{\alpha\beta}{}^\gamma T_\gamma$.

Representation matrices in some representation (and its dual) labeled by indices $i,j,\ldots$ are denoted $t_{\alpha i}{}^j$, and fulfill 
\begin{align}
[t_\alpha,t_\beta]\equiv
t_\alpha t_\beta-(-1)^{\alpha\beta}t_\beta t_\alpha=f_{\alpha\beta}{}^\gamma t_\gamma\;.
\label{RepMEq}
\end{align}
Representation matrices in the adjoint representation are
$t_{\alpha\beta}{}^\gamma=f_{\beta\alpha}{}^\gamma$, which is verified using the graded Jacobi identity:
\begin{align}
[t_\alpha,t_\beta]_\gamma{}^\delta
&=f_{\gamma\alpha}{}^\epsilon f_{\epsilon\beta}{}^\delta
-(-1)^{\alpha\beta}f_{\gamma\beta}{}^\epsilon f_{\epsilon\alpha}{}^\delta 
=-(-1)^{(\alpha+\beta)\gamma}f_{\alpha\beta}{}^\epsilon f_{\epsilon\gamma}{}^\delta \nn\\
&=(-1)^{(\alpha+\beta+\epsilon)\gamma}f_{\alpha\beta}{}^\epsilon f_{\gamma\epsilon}{}^\delta 
=f_{\alpha\beta}{}^\epsilon t_{\epsilon\gamma}{}^\delta\;.
\end{align} 

To maintain a tensor formalism, multiplying objects (tensors in graded modules), one needs to keep track of the order in which they are multiplied. We use the Munich conventions, where index contractions are always done on nearby indices in the NW-SE direction, so an expression $U^iV_i$ becomes a scalar. The action of $A=A^\alpha T_\alpha\in\fg$ is defined as $(A\cdot U)^i=-U^jA^\alpha t_{\alpha j}{}^i$, $(A\cdot V)_i=A^\alpha t_{\alpha i}{}^jV_j$.
For covariant derivatives, we use
$DX\equiv\dot X-A\cdot X$.

In order for matrices to be multiplied with each other, an index structure $M_i{}^j$ is necessary. Such matrices, when non-singular, have inverses.
For tensors with index structure $m_{ij}$ one must use a super-inverse $\tilde m$, defined by 
$\tilde m^{ik}m_{kj}=(-1)^i\delta^i{}_j$, which is a tensor since transposition moves the two indices past each other. This applies \eg\ to the symplectic structure $\omega$ used in Section \ref{KTBVSec}, and is the same sign responsible for super-trace $\Str M=(-1)^iM_i{}^i$.

Objects, for example ghosts, may appear in parity-shifted modules. We then have, for adjoint ghosts, $|c^\alpha|=|\alpha|+1$, so 
$c^\alpha c^\beta=(-1)^{(\alpha+1)(\beta+1)}c_\beta c_\alpha$. One might worry that there is a clash between this {\it parity-shifted graded symmetry} and the {\it graded antisymmetry} of the structure constants. They differ in that the former implies ``$\hbox{even}\times\hbox{odd}=\hbox{odd}\times\hbox{even}$'', while the latter gives
``$\hbox{even}\times\hbox{odd}=-\hbox{odd}\times\hbox{even}$''. The two different symmetrisations are equivalent (for example as modules of a general linear superalgebra), see \eg\ ref. \cite{Cederwall:2018aab}, and can be translated into each other. This translation is taken care of ``automatically''.
BRST, or BV, variations of various objects
typically contain $c^\alpha t_{\alpha i}{}^j$. Matrix multiplication of this object with itself would implement parity-shifted graded symmetry on $t_\alpha t_\beta$, which is not good.  
Instead, a sign twist must appear (and indeed always does) so the object appearing is 
$C_i{}^j\equiv c^\alpha\tilde t_{\alpha i}{}^j$, where
$\tilde t_{\alpha i}{}^j=(-1)^it_{\alpha i}{}^j=(-1)^{\alpha+j}t_{\alpha i}{}^j$. 
The $\tilde t$'s obey
\begin{align}
(\tilde t_\alpha \tilde t_\beta+(-1)^{(\alpha+1)(\beta+1)}
\tilde t_\beta\tilde t_\alpha)_i{}^j
=(-1)^\alpha f_{\alpha\beta}{}^\gamma t_{\alpha i}{}^j\;,
\end{align}
so they are naturally not graded commuted, but multiplied with parity-shifted graded symmetry.
Another way to observe the necessity of this sign is to 
note that 
\begin{align}
U^iC_i{}^j=(-1)^iU^ic^\alpha t_{\alpha i}{}^j
=(-1)^{i+i(\alpha+1)}c^\alpha U^it_{\alpha i}{}^j
=(-1)^{i\alpha}c^\alpha U^it_{\alpha i}{}^j\;,
\end{align}
is a tensor: the sign $(-1)^{i\alpha}$ compensates for the ${}^i{}_i$ contraction skipping an $\alpha$. Alternatively, the sign $(-1)^i$ is attributed to $U^i$ skipping an odd contraction $c^\alpha t_\alpha$.
Thus, also $C_i{}^j$ is a tensor.

One may note that the structure constants $f_{\alpha\beta}{}^\gamma$ are less natural than the representation matrices (and it would indeed have been better to define them with lower indices in the opposite order). Namely, 
eq. \eqref{RepMEq} is not explicitly tensorial.
Define the tensor 
$Y_{\alpha\beta i}{}^j
=(-1)^{\alpha\beta}t_{\alpha i}{}^k t_{\beta k}{}^j$, where the sign factor comes from 
taking $\beta$ past $k$ and $i$ ($(-1)^{(i+k)\beta}=(-1)^{\alpha\beta}$). Then eq. \eqref{RepMEq} becomes
\begin{align}
-2Y_{[\alpha\beta]i}{}^j=f_{\beta\alpha}{}^\gamma t_{\gamma i}{}^j
=t_{\alpha\beta}{}^\gamma t_{\gamma i}{}^j\;,
\end{align}
which is manifestly tensorial, relating the first and last expressions with the same index structure.
Many similar sign manipulations are behind the calculations in Section \ref{KTBVSec}.

\bibliographystyle{utphysmod2}

\bibliography{biblio}

\end{document}